\documentclass[12pt]{article}
\usepackage{epsfig}
\usepackage{multicol}
\usepackage{graphicx, slashed, amsmath, amsfonts, amssymb}
\usepackage[onehalfspacing]{setspace}
\usepackage[top=1in, bottom=1in, left=1in, right=1in]{geometry}
\usepackage{rotating}
\usepackage{subcaption}
\usepackage{cite}
\def\bc{\begin{center}}
\def\ec{\end{center}}
\def\be{\begin{equation}}
\def\ee{\end{equation}}
\def\ba{\begin{array}}
\def\ea{\end{array}}
\def\beqn{\begin{eqnarray}}
\def\eeqn{\end{eqnarray}}
\begin{document}
\title{Probing texture 4 zero quark mass matrices in the era of precision measurements}
\author{Aakriti Bagai, Aseem Vashisht, Nikhila Awasthi,\\  Gulsheen Ahuja$^{*}$, Manmohan Gupta$^{+}$\\
{\it Department of Physics, Centre of Advanced Study,}\\
{\it Panjab University, Chandigarh, India.}\\
*gulsheen@pu.ac.in, $^{+}$mmgupta@pu.ac.in
{}}\maketitle

\begin{abstract}
In pursuance of finding a viable set of mass matrices, keeping in mind recent precision measurement of quark masses and CKM parameters, we have carried out a detailed and comprehensive  analysis of hermitian texture 4 zero quark mass matrices. Our analysis reveals that only Fritzsch like texture 4 zero combination provides a viable set of mass matrices for the present quark mixing data and all other texture 4 zero combinations are ruled out. A good number of these possibilities can be ruled out analytically, the other possibilities are excluded by the present quark mixing data. Interestingly, our conclusions remain valid, even if in future there are perturbations in the ranges of the light quark masses or CKM parameters.

\end{abstract}
\section{Introduction}
Understanding the fermion masses and mixings, usually referred to as Flavor Physics, is one of the major goals of current research in present day High Energy Physics. Understanding it from more fundamental considerations poses a big challenge for both experimentalists and theoriticians, therefore, making this area of research to be a fertile one. On the experimental front, in the last few years, the main strategy has been to sharpen the tools of observation, providing remarkable progress in the measurements of the fermion masses and mixing parameters. In particular, for the quark sector, the Cabibbo Kobayashi Maskawa (CKM) parameters \cite {Cabibbo, KM} can now be considered to be known at the level of `precision measurements' in the context of CKM phenomenology \cite{pdg20}. Similarly, considerable progress has been made in the measurement of the quark masses, in particular of the light quark masses, $m_u$, $m_d$ and $m_s$ \cite{masses}. In view of the relationship of the CKM matrix with the quark mass matrices, these developments regarding measurements of the quark masses as well as the CKM parameters would undoubtedly have deep implications for the structure of the mass matrices. 

The theoretical understanding of fermion masses and mixings proceeds along two approaches, i.e., ‘top-down’ \cite{gu}- \cite{superstrings} and ‘bottom-up’ \cite{bu1}-\cite{bu5} . The essential idea behind `bottom-up' approach is that one tries to find the phenomenological fermion mass matrices which are in tune with the low energy data and can serve as guiding stone for developing more ambitious theories. It may be noted that in the SM, the fermion mass matrices are arbitrary, i.e., these constitute a set of two 3 × 3 general complex mass matrices with 36 real free parameters. This large number of free parameters has to account for a much smaller number of physical observables, e.g., in the quark sector, the mass matrices, $M_U$ and $M_D$, need to describe 10 physical observables, i.e., 6 non vanishing quark masses, 3 mixing angles and 1 CP violating phase in the standard parametrization of the CKM matrix. Similarly, in the leptonic sector, physical observables described by lepton mass matrices are 6 lepton masses, 3 mixing angles and 1 CP violating phase for Dirac neutrinos (two additional phases in case neutrinos are Majorana particles). Therefore, as a first step, in order to develop viable phenomenological fermion mass matrices one has to limit the number of free parameters in the mass matrices.

In this context, it is well known that in the SM and its extensions a rotation of the right-handed fermion fields does not affect any physical results, hence, the mass matrices can be considered as hermitian without loss of generality. This immediately brings down the number of real free parameters from 36 to 18, which however, is still a large number compared to the number of observables. To this end, Weinberg \cite{weinberg} implicitly and Fritzsch \cite{fri1,fri2} explicitly initiated the idea of texture zero mass matrices wherein some of the entries of the mass matrices were proposed to be zero. As a result, a fewer number of free parameters imparted more predictability to mass matrices. A particular texture `n' zero mass matrix is defined such that sum of the number of zeros at diagonal positions and a pair of symmetrically placed zeros at off-diagonal positions, counted as one, is `n'.   The original Fritzsch ansatz of quark mass matrices is given by \cite{fri1,fri2}

 \begin{equation}
    M_{U}=\begin{pmatrix}
      0 & A_{U} & 0 \\
      A^{*}_{U} & 0 & B_{U} \\
      0 & B_{U}^{*} & C_{U} \\
    \end{pmatrix} ,\quad M_{D}=\begin{pmatrix}
      0 & A_{D} & 0 \\
      A^{*}_{D} & 0 & B_{D} \\
      0 & B_{D}^{*} & C_{D} \\
    \end{pmatrix},
\label{mumd}
\end{equation}
where $M_{U}$ and $M_{D}$ correspond to mass matrices in the up (U) and down (D) sector with complex off diagonal elements, i.e., $A_{i}=|A_{i}|e^{\text{i$\alpha $}}$ and $B_{i}=|B_{i}|e^{\text{i$\beta $}}$, where $i =U,D $, whereas $C_{i}$ is the real element of the matrix. 
Both the matrices $M_U$ and $M_D$ are texture 3 zero type, together this set of matrices is referred to as texture 6 zero mass matrices. This combination is also called minimal texture structure since it has 10 free parameters i.e., equal to the number of physical observables in the quark sector.

The methodology of examining the viability of a particular texture zero combination involves arriving at the corresponding CKM matrix, the compatibility of which is examined with respect to the experimentally obtained one, given by Particle Data Group (PDG), concluding whether the texture combination being analyzed is a viable one. Several analyses of the above mentioned combination of texture 6 zero matrices, given in equation \eqref{mumd}, were carried out \cite{tex6/5zero1}-\cite{tex6/5zero2} revealing these matrices to be non viable as the CKM matrix constructed using these was found to be incompatible with the one obtained experimentally. Very recently, a detailed and comprehensive analysis of not only these mass matrices but also of all possible non-Fritzsch texture 6 zero mass matrices has been carried out \cite{6zero}. It has been shown that all these combinations are ruled out either analytically or by the present quark mixing data. 

An immediate extension of the above mentioned minimal texture is the texture 5 zero mass matrices which can be obtained by replacing one of the zero entry in one of the texture 3 zero matrix with a non zero one. For example, considering the (2,2) element of any of the above mentioned matrices $M_U$ and $M_D$ to be non zero, one obtains the following texture 2 zero mass matrix
 \begin{equation}
    M_{i}=\begin{pmatrix}
      0 & A_{i} & 0 \\
      A^{*}_{i} & D_{i} & B_{i} \\
      0 & B_{i}^{*} & C_{i} \\
    \end{pmatrix},
\label{mumd2}
\end{equation}
where $i=U, D$ and $D_i$ is the real element of the matrix. Along with these matrices, depending upon the position of zeros there are several other possible structures which can be considered to be texture 2 zero ones. Thereafter, one arrives at texture 5 zero mass matrices by considering either of the mass matrix in the up or the down sector to be texture 3 zero type, e.g., the one given in equation \eqref{mumd}, along with the mass matrix in the other sector being 2 zero type, e.g., the one given in equation \eqref{mumd2}. Several detailed analyses of this texture of quark mass matrices have also been carried out in the literature \cite{tex6/5zero1,tex6/5zero2}. In a very recent work \cite{5zero}, it has been shown that all these possibilities are now excluded by the present quark mixing data.
 
As a next step, we discuss the texture 4 zero combinations which can be obtained when one considers both $M_U$ and $M_D$ to be texture 2 zero type. In case, both $M_U$ and $M_D$ have the structure mentioned in equation \eqref{mumd2}, the combination is said to be Fritzsch-like texture 4 zero quark mass matrices. Interestingly, this texture has been analyzed earlier \cite{paraspace1, paraspace2}, wherein the authors have examined the compatibility of Fritzsch-like as well as some other possibilities of texture 4 zero mass matrices. 
Further, in Ref. \cite{unique}, beginning with the most `general' mass matrices within the context of SM, and imposing Weak Basis Transformations \cite{bu1, wbt} as well as the condition of `naturalness' \cite{natural},  texture 4 zero mass matrix was obtained and shown to be a unique viable option for the description of quark mixing data. It may also be mentioned that the Fritzsch like texture 4 zero mass matrices are known to be compatible with specific models of GUTs, e.g., SO(10) \cite{1so10}-\cite{3so10} and these could be obtained using considerations of Abelian family symmetries \cite{abelian}.  

Keeping in mind the above mentioned issues and considering refinements in measurement of quark masses \cite{masses, ratios1, ratios2}, it becomes desirable to not only revisit Fritzsch like texture 4 zero mass matrices mentioned in equation \eqref{mumd2}, but also, to carry out a detailed and comprehensive analysis of a large number of possible texture 4 zero combinations. The purpose of the present work is to first enumerate all possible texture 4 zero mass matrices. As a next step, a detailed analysis pertaining to the viability of all these possible combinations of the mass matrices has been carried out with an emphasis to examine if, in future, there are changes in the ranges of the light quark masses, whether or not, the conclusions remain same. The detailed plan of the paper is as follows. In Section (2), we enlist the various possibilities of texture 2 zero matrices and detail the essentials of the formalism regarding the texture four zero mass matrices. Inputs used in the present analysis have been given in Section (3) and the discussion of the calculations and results have been presented in Section (4). Finally, Section (5) summarizes our conclusions.
\section{Permutation based enumeration of texture 4 zero combinations}

To obtain the total number of texture 2 zero structures, one can make use of the fact that the total number of structures for a texture `n' zero mass matrix is $^6C_n = \frac{6!}{n!(6-n)!}, $ 6 being the number of ways to enter zeros in the mass matrices. For $n=2$, one gets the following 15 possible structures, ${S_1}$ to $S_{15}$, for texture 2 zero mass matrices:

 \begin{list}{(i)}
  \item One zero along diagonal position and two zeros symmetrically placed at off diagonal positions:\\
 \begin{center}
	$S_{1}= \begin{pmatrix}
	0	& \times & 0 \\ 
	\times	& \times  &  \times\\ 
	0	& \times  & \times
	\end{pmatrix},\: S_{2}= \begin{pmatrix}
	0	& 0 & \times \\ 
	0	&\times  &  \times\\ 
	\times	&\times  & \times
\end{pmatrix},\:
S_{3}= \begin{pmatrix}
\times	& \times & \times \\ 
\times	& 0  & 0\\ 
\times	& 0  & \times
\end{pmatrix},$\\
$S_{4}= \begin{pmatrix}
\times	& \times & 0 \\ 
\times	&\times  &\times\\ 
0	& \times  & 0
\end{pmatrix},\:S_{5}= \begin{pmatrix}
\times	& \times & \times \\ 
\times	& \times  & 0\\ 
\times	& 0  & 0
\end{pmatrix},\: S_{6}= \begin{pmatrix}
\times	& 0 & \times \\ 
0	& 0  & \times\\ 
\times	& \times  & \times
\end{pmatrix},$
$S_{7}= \begin{pmatrix}
\times	& \times & 0 \\ 
\times	&0  & \times\\ 
0	& \times  & \times
\end{pmatrix},\: S_{8}= \begin{pmatrix}
\times	& 0 & \times \\ 
0	& \times  & \times\\ 
\times	& \times & 0
\end{pmatrix}, \:S_{9}= \begin{pmatrix}
0	& \times & \times \\ 
\times	& \times  & 0\\ 
\times	& 0  & \times
\end{pmatrix}.$
\end{center}
\end{list}

 \begin{list}{(ii)}
	\item Two zeros along diagonal positions:\\	
\begin{center}
$S_{10}= \begin{pmatrix}
	0	& \times & \times \\ 
	\times	&0  &  \times\\ 
	\times	&\times  & \times
	\end{pmatrix},\: S_{11}= \begin{pmatrix}
	0	& \times & \times \\ 
	\times	&\times  &  \times\\ 
	\times	&\times  & 0
\end{pmatrix},\:
S_{12}= \begin{pmatrix}
\times	& \times & \times \\ 
\times	& 0  & \times\\ 
\times	& \times  & 0
\end{pmatrix},$\\ 
\end{center}	
where $\times$'s represent the non-vanishing entries.
\end{list}

\begin{list}{(iii)}
	\item Two zeros symmetrically placed at two off diagonal positions:\\
	\begin{center}
		$ S_{13}= \begin{pmatrix}
		\times	& \times & 0 \\ 
			\times	&\times  &  0\\ 
			0	& 0  & \times
		\end{pmatrix},\: S_{14}= \begin{pmatrix}
			\times	& 0 & \times \\ 
			0	& \times &  0\\ 
			\times	& 0  & \times
		\end{pmatrix},\:
		S_{15}= \begin{pmatrix}
			\times	& 0 & 0 \\ 
			0	& \times  & \times\\ 
			0	& \times  & \times
		\end{pmatrix}.$\\
		
	\end{center}
	\end{list}
   
In general, one has the freedom to consider the mass matrices in the up and down sectors, i.e., $M_{U}$ and $M_{D}$ to be either  of the above listed 15 patterns, resulting into $15 \times 15 = $ 225 combinations corresponding to texture 4 zero mass matrices.
 
It may be noted that the structure $S_1$ correspond to the Fritzsch-like \emph{ans$\ddot{a}$tz} mentioned in equation (\ref{mumd2}). Interestingly, one finds that the structures $S_1$, $S_2$, $S_3$, $S_4$, $S_{5}$ and $S_{6}$ are related as
\be
S_j = p_j^T S_1 p_j, ~~~~~~~~(j=1-6)
\label{permutation}
\ee
where $p_j$ are the following 6 permutation matrices
\begin{equation*}
p_1=\left(
\begin{array}{ccc}
	1 & 0 & 0 \\
	0 & 1 & 0 \\
	0 & 0 & 1
\end{array}
\right) ,\: p_2=\left(
\begin{array}{ccc}
	1 & 0 & 0 \\
	0 & 0 & 1 \\
	0 & 1 & 0
\end{array}
\right),\: p_3=\left(
\begin{array}{ccc}
	0 & 1 & 0 \\
	1 & 0 & 0 \\
	0 & 0 & 1
\end{array}
\right),
\end{equation*}	\\
\begin{equation}
	p_4=\left(
	\begin{array}{ccc}
		0 & 0 & 1 \\
		0 & 1 & 0 \\
		1 & 0 & 0
	\end{array}\right),\:
	p_5=\left(
	\begin{array}{ccc}
		0 & 0 & 1 \\
		1 & 0 & 0 \\
		0 & 1 & 0
	\end{array}
	\right),\: p_6=\left(
	\begin{array}{ccc}
		0 & 1 & 0 \\
		0 & 0 & 1 \\
		1 & 0 & 0
	\end{array}
	\right).
\label{permutations}	
\end{equation}	

These 6 matrices, $S_{1}$ to $S_{6}$, have been placed in Class I of Table \ref{table1} and for further discussion would be referred as I$_a$, I$_b$, etc..

It should be kept in mind that the Fritzsch like texture 2 zero matrix given in equation (\ref{mumd2}) has been obtained from texture 3 zero Fritzsch \emph{ans$\ddot{a}$tz}, given in equation (\ref{mumd}), by replacing the zero at $(2,2)$ with a non zero element. However, it may be mentioned again that there are other ways as well to arrive at texture 2 zero matrices from texture 3 zero ones, e.g., one can firstly replace the zero entry at $(1,1)$ in Fritzsch matrix given in equation (\ref{mumd}) with a non zero one, secondly  both the zeros can be placed at diagonal positions with all other elements being non-zero, and thirdly both the zero entries can symmetrically be placed at off-diagonal positions. The matrices thus obtained, have been placed as the first matrix in Class II, Class III and Class IV of table \ref{table1} respectively.  We then use the permutation relation, mentioned in equation (\ref{permutation}), to obtain 5 more matrices of the respective classes, i.e., matrices II$_b$ - II$_f$,  III$_b$ - III$_f$ and IV$_b$ - IV$_f$. It is interesting to note that the matrices II$_a$, II$_b$ and II$_c$ have structures $S_7$, $S_8$ and $S_9$ respectively. The remaining three matrices of this class, namely II$_d$, II$_e$ and II$_f$ also have the same structures as the other three, i.e., these correspond to structures $S_8$, $S_9$ and $S_7$ respectively. Hence, we can say that out of 6 matrices belonging to Class II, only three are structurally different. Similar to the Class II matrices, the three matrices of Classes III and IV also have same structures as the other three in their respective class. One may note that both III$_a$ and III$_c$ have structure $S_{10}$, III$_b$ and III$_e$ have structure $S_{11}$, while III$_d$ and III$_f$ correspond to structure $S_{12}$. Also, three matrices, grouped in Class IV, i.e., IV$_a$, IV$_b$ and IV$_d$ have structures $S_{13}$, $S_{14}$ and $S_{15}$ respectively, which are also the corresponding structures of the other three matrices of this class, namely, IV$_c$, IV$_e$ and IV$_f$. Interestingly, the matrices having similar structures yield same results, implying exactly the same corresponding CKM matrix. Therefore, these corresponding matrices should not be counted twice implying that in all, one obtains 15 independent structures for $M_{U}$ and $M_{D}$ being texture 2 zero type, 6 belonging to Class I and 3 each to Class II, III and IV of the table.

\begin{table}

\scriptsize

\begin{tabular}{|c|c|c|c|c|}

	\hline 
	&\text{Class I} &\text{Class II} & \text{Class III} & \text{Class IV}  \\  
	\hline
a	& $\left(
\begin{array}{ccc}
	0 & A e^{\text{i$\alpha $}} & 0 \\
	A e^{-\text{i$\alpha $}} & D & B e^{\text{i$\beta $}} \\ 
	0 & B e^{-\text{i$\beta $}} & C
\end{array}
\right)$ & $\left(
\begin{array}{ccc}
	E & A e^{\text{i$\alpha $}} & 0 \\
	A e^{-\text{i$\alpha $}} & 0 & B e^{\text{i$\beta $}} \\
	0 & B e^{-\text{i$\beta $}} & C
\end{array}
\right)$  & $\left(
\begin{array}{ccc}
	0 & A e^{\text{i$\alpha $}} & F e^{\text{i$\gamma $}} \\
	A e^{-\text{i$\alpha $}} & 0 & B e^{\text{i$\beta $}} \\
	F e^{-\text{i$\gamma $}} & B e^{-\text{i$\beta $}} & C
\end{array}
\right)$  & $\left(
\begin{array}{ccc}
	E & A e^{\text{i$\alpha $}} & 0 \\
	A e^{-\text{i$\alpha $}} & D & 0 \\
	0 & 0 & C
\end{array}
\right)$ \\ 
\hline 
b	& $\left(
\begin{array}{ccc}
0 & 0 & A e^{\text{i$\alpha $}} \\
0 & C & B e^{-\text{i$\beta $}} \\
A e^{-\text{i$\alpha $}} & B e^{\text{i$\beta $}} & D
\end{array}
\right)$ & $\left(
\begin{array}{ccc}
	E & 0 &  A e^{\text{i$\alpha $}} \\
	0 & C & B e^{-\text{i$\beta $}} \\
	 A e^{-\text{i$\alpha $}} & B e^{\text{i$\beta $}} & 0
\end{array}
\right)$  & $\left(
\begin{array}{ccc}
	0 & F e^{\text{i$\gamma $}} & A e^{\text{i$\alpha $}} \\
	F e^{-\text{i$\gamma $}} & C & B e^{-\text{i$\beta $}} \\
	A e^{-\text{i$\alpha $}} & B e^{\text{i$\beta $}} & 0
\end{array}
\right)$ & $\left(
\begin{array}{ccc}
E & 0 & A e^{\text{i$\alpha $}} \\
0 & C & 0 \\
A e^{-\text{i$\alpha $}} & 0 & D
\end{array}
\right)$ \\ 
	\hline 
c	& $\left(
\begin{array}{ccc}
D & A e^{-\text{i$\alpha $}} & B e^{\text{i$\beta $}} \\
A e^{\text{i$\alpha $}} & 0 & 0 \\
B e^{-\text{i$\beta $}} & 0 & C
\end{array}
\right)$ & $\left(
\begin{array}{ccc}
	0 &  A e^{-\text{i$\alpha $}} & B e^{\text{i$\beta $}} \\
	 A e^{\text{i$\alpha $}} & E & 0 \\
	 B e^{-\text{i$\beta $}} & 0 & C
\end{array}
\right)$  & $\left(
\begin{array}{ccc}
	0 & A e^{-\text{i$\alpha $}} & B e^{\text{i$\beta $}} \\
	A e^{\text{i$\alpha $}} & 0 & F e^{\text{i$\gamma $}} \\
	B e^{-\text{i$\beta $}} & F e^{-\text{i$\gamma $}} & C
\end{array}
\right)$ & $\left(
\begin{array}{ccc}
D & A e^{-\text{i$\alpha $}} & 0 \\
A e^{\text{i$\alpha $}} & E & 0 \\
0 & 0 & C
\end{array}
\right)$ \\ 
	\hline 
d	& $\left(
\begin{array}{ccc}
C & B e^{-\text{i$\beta $}} & 0 \\
B e^{\text{i$\beta $}} & D & A e^{-\text{i$\alpha $}} \\
0 & A e^{\text{i$\alpha $}} & 0
\end{array}
\right)$  & $\left(
\begin{array}{ccc}
	C & B e^{-\text{i$\beta $}} & 0 \\
	B e^{\text{i$\beta $}} & 0 & A e^{-\text{i$\alpha $}}\\
	0 & A e^{\text{i$\alpha $}} & E
\end{array}
\right)$  & $\left(
\begin{array}{ccc}
	C & B e^{-\text{i$\beta $}} & F e^{-\text{i$\gamma $}} \\
	B e^{\text{i$\beta $}}& 0 & A e^{-\text{i$\alpha $}}  \\
	F e^{\text{i$\gamma $}} &  A e^{\text{i$\alpha $}} & 0
\end{array}
\right)$ &   $\left(
\begin{array}{ccc}
C & 0 & 0 \\
0 & D & A e^{-\text{i$\alpha $}} \\
0 & A e^{\text{i$\alpha $}} & E
\end{array}
\right)$  \\ 
	\hline 
e	& $\left(
\begin{array}{ccc}
D & B e^{\text{i$\beta $}} & A e^{-\text{i$\alpha $}} \\
B e^{-\text{i$\beta $}} & C & 0 \\
A e^{\text{i$\alpha $}} & 0 & 0
\end{array}
\right)$ & $\left(
\begin{array}{ccc}
	0 & B e^{\text{i$\beta $}} & A e^{-\text{i$\alpha $}} \\
	B e^{-\text{i$\beta $}} & C & 0 \\
	A e^{\text{i$\alpha $}} & 0 & E
\end{array}
\right)$  & $\left(
\begin{array}{ccc}
	0 & B e^{\text{i$\beta $}} & A e^{-\text{i$\alpha $}} \\
	B e^{-\text{i$\beta $}} & C & F e^{-\text{i$\gamma $}} \\
	 A e^{\text{i$\alpha $}} & F e^{-\text{i$\gamma $}} & 0
\end{array}
\right)$ & $\left(
\begin{array}{ccc}
D & 0 & A e^{-\text{i$\alpha $}} \\
0 & C & 0 \\
A e^{\text{i$\alpha $}} & 0 & E
\end{array}
\right)$ \\ 
	\hline 
f	& $\left(
\begin{array}{ccc}
C & 0 & B e^{-\text{i$\beta $}} \\
0 & 0 & A e^{\text{i$\alpha $}} \\
B e^{\text{i$\beta $}} & A e^{-\text{i$\alpha $}} & D
\end{array}
\right) $ & $\left(
\begin{array}{ccc}
	C & 0 & B e^{-\text{i$\beta $}} \\
	0 & E & A e^{\text{i$\alpha $}} \\
	B e^{\text{i$\beta $}} & A e^{-\text{i$\alpha $}} & 0
\end{array}
\right)$  & $\left(
\begin{array}{ccc}
	C & F e^{-\text{i$\gamma $}} & B e^{-\text{i$\beta $}} \\
	F e^{\text{i$\gamma $}} & 0 & A e^{\text{i$\alpha $}} \\
	B e^{\text{i$\beta $}} & A e^{-\text{i$\alpha $}} & 0
\end{array}
\right)$ & $\left(
\begin{array}{ccc}
C & 0 & 0 \\
0 & E & A e^{\text{i$\alpha $}} \\
0 & A e^{-\text{i$\alpha $}} & D
\end{array}
\right)$ \\ 
	\hline 
\end{tabular}
\vspace{20pt}
\caption{Possible texture 2 zero mass matrices belonging to Class I, II, III and IV} 
\label{table1} 
\end{table}

Following the methodology presented in Ref. \cite{6zero} and \cite{5zero}, one may note that it essentially
involves considering a possible texture 4 zero combination, i.e., $M_{U}$ and $M_{D}$ being any of the texture 2 zero type, i.e., 15 independent patterns listed in Classes I-IV of table \ref{table1}. The viability of the considered combination is explored by examining the compatibility of the CKM matrix constructed from a given combination of mass matrices, with the recent one given by PDG. The standard structure of the CKM matrix given by PDG is 
\begin{equation}
V_{CKM} =\begin{pmatrix}
c_{12}c_{13} & s_{12}c_{13} & s_{13}e^{-i\delta} \\ 
-s_{12}c_{23}-c_{12}s_{23}s_{13}e^{i\delta} & c_{12}c_{23}-s_{12}s_{23}s_{13}e^{i\delta} & s_{23}c_{13} \\ 
s_{12}s_{23}-c_{12}c_{23}s_{13}e^{i\delta} & -c_{12}s_{23}-s_{12}c_{23}s_{13}e^{i\delta} & c_{23}c_{13}
\end{pmatrix} ,
\label{standard}
\end{equation}
where $ c_{ij}=\cos \theta_{ij} $ and $ s_{ij}=\sin \theta_{ij} $. 
The latest values of various parameters of the matrix, given by PDG 2020 \cite{pdg20}, are as following 
\begin{equation}
\sin \theta_{12} = 0.22650 \pm 0.00048,~~~\sin\theta_{13} = 0.00361^{+0.00011}_{-0.00009},~~~\sin\theta_{23} = 0.04053^{+0.00083}_{-0.00061},~~~\delta=1.196^{+0.045}_{-0.043}.
\end{equation} 
It may be noted that substituting these values in matrix given in equation (\ref{standard}) leads to values of the diagonal elements nearly equal to unity, while the off diagonal elements have much smaller values. Thus, the standard structure of CKM matrix is where the diagonal elements are of the order 1, while the off diagonal elements are much smaller.

To construct CKM matrix from the mass matrices,  one needs to find the diagonalizing transformations of the corresponding mass matrices. For this purpose, the real matrix corresponding to $M_{i}$ ($i=U, D$) can be expressed as
  \begin{equation}    
   M_{i}=P_{i}^{\dag}M_{i}^{r}P_{i}, 
  \end{equation}
where  $ M_{i}^{r}$ is real matrix and $P_{i}$ denotes the phase matrix. An essential step for the construction of the diagonalizing transformation matrix is to consider the invariants, trace $M_{i}^r$, trace $M_{i}^{r^{2}}$ and determinant $M_{i}^r$ which yield relations involving elements of mass matrices.
For all the 6 matrices belonging to Class I, we obtain the following relations:
\begin{equation}
C_{i}+D_{i}=m_{1}-m_{2}+m_{3},~~~~|A_{i}|^2+|B_{i}|^2-C_{i}D_{i}=m_1m_2+m_2m_3-m_1m_3,~~~~|A_{i}|^2C_{i}=m_1m_2m_3.
\label{eqIa}
\end{equation}
Similarly, for the Class II matrices, diagonalization equations are
\begin{equation}
C_{i}+E_{i}=m_{1}-m_{2}+m_{3},~~~~2|A_{i}|^2+2|B_{i}|^2-C_{i}E_{i}=m_1m_2+m_2m_3-m_1m_3,~~~~|A_{i}|^2C_{i}+|B_{i}|^2E_{i}=m_1m_2m_3.
\label{class2}
\end{equation}
The Class III matrices satisfy the following equations
\begin{equation}
C_{i}=m_{1}-m_{2}+m_{3},~~~~|A_{i}|^2+|B_{i}|^2+|F_{i}|^2=m_1m_2+m_2m_3-m_1m_3,~~~~|A_{i}|^2C_{i}-2|A_{i}||F_{i}||B_{i}|=m_1m_2m_3.
\label{class3}
\end{equation}
For the matrices placed in Class IV, the diagonalization equations are
\begin{equation}
\begin{aligned}
C_{i}+D_{i}+E_{i}=m_{1}-m_{2}+m_{3},\\
|A_{i}|^2-E_{i}D_{i}-E_{i}C_{i}-D_{i}C_{i}=m_1m_2+m_2m_3-m_1m_3, \\
|A_{i}|^2C_{i}-E_{i}D_{i}C_{i}=m_1m_2m_3.
\end{aligned}
\label{class4}
\end{equation}
The real matrix $M_{i}^{r}$ can then be diagonalized by the orthogonal transformations $O_{i}$, i.e.,
\begin{equation}
  M_{i}^{diag}= O_{i}^{T}M_{i}^{r}O_{i}=
  O_{i}^{T}P_{i}M_{i}P_{i}^{\dag}O_{i}= U_i^{\dagger}M_{i}U_i,
  \label{diagonalize}
  \end{equation}
where $ O_{i}^{T}P_{i} = U_i^{\dagger}$ and $P_{i}^{\dag}O_{i}= U_i$ denotes diagonalizing unitary matrices. $M_{i}^{diag}={\rm diag}(m_{1},-m_{2},m_{3})$, where the subscripts 1, 2 and 3 refer respectively to u, c and t for the up sector and d, s and b for the down sector.

It has been shown in an earlier analysis \cite{Cl31,Cl32} that exact diagonalization of Class III matrices is not possible. This can also be seen from the diagonalizing equations for Class III matrices given in equation (\ref{class3}), hence, one cannot construct CKM matrix from these matrices. Also, the relations for Class IV matrices, given in equation (\ref{class4}), lead to decoupling of a generation of quarks \cite{Cl4}, i.e., the matrices placed in class IV lead to two family mixing only, hence, all the matrices placed in Class III and IV are ruled out. 

As a next step of our analysis, we present all possible texture 4 zero combinations, wherein, $M_U$ and $M_D$ can be considered from Class I and/or Class II of the table, leading to the following:
\\
Category 1: $M_{U}$ and $M_{D}$ both from Class I, resulting into $6 \times 6 = 36 $ combinations.\\
Category 2: $M_{U}$ and $M_{D}$ both from Class II, resulting in $3 \times 3 = 9 $ possible set of mass matrices. \\
Category 3: $M_{U}$ from Class I and $M_{D}$ from Class II. This results in $6 \times 3 = 18$ possible combinations\\
Category 4: $M_{U}$ from Class II and $M_{D}$ from Class I, leading to a total of $3 \times 6 =18$ combinations.\\ 
 
Thus, one needs to examine 36+9+18+18 = 81 texture 4 zero combinations. However, it may be noted that since the matrices belonging to each class are related through permutation relations, the CKM matrices obtained from these are also related.
As mentioned earlier, equation (\ref{permutation}) shows that all the matrices belonging to a particular class are related to the first matrix of the class through permutation matrices, i.e.,
\begin{equation}
M_i = p_j^T M_i^{a} p_j,
\end{equation}
where $M_i^{a}$ denotes the first matrix belonging to a class. Therefore, the equation (\ref{diagonalize}) can be re-written as 
\begin{equation}
M_i^{diag} = U_i^{\dagger} p_{j}^T M_i^{a} p_{j} U_i
\label{diagonalizea}
\end{equation} 
If the diagonalising matrix for $M_i^{a}$ is denoted by $U_i^a$, then using the relation given in equation (\ref{diagonalize}), one can write 
\begin{equation}
M_i^{diag} = U_i^{a \dagger} M_i^{a} U_i^a
\label{diagonaliza}
\end{equation}
On comparing the equations (\ref{diagonalizea}) and (\ref{diagonaliza}), it can be seen that
\begin{equation}
 U_i^{a \dagger}=U_i^{\dagger}  p_j^T ~~~~ and ~~~~U_i^a= p_j U_i
\end{equation}
or using the property $p_j p_j^T = p_j^T p_j = I$, where $I$ is the identity matrix, the above equation can be re-written as
\begin{equation}
 U_i^{\dagger} = U_i^{a \dagger} p_j ~~~~ and ~~~~U_i= p_j^T U_i^a
\end{equation}
The CKM matrix can then be obtained as 
\begin{equation}
V_{CKM}= U_U^{\dagger} U_D = U_U^{a \dagger} p_j p_{j'}^T U_D^{a}
\label{vckm}
\end{equation} 
where $p_j$ and $p_{j'}$ respectively denotes the permutation used to obtain $M_U$ and $M_D$ from the first matrices ($M_U^a$ and $M_D^a$) of their respective class. 
It is interesting to note that since the product of two permutation matrices gives another permutation matrix, i.e., 
\begin{equation}
p_j p_{j'}^T = p_{k}~~~~~~~~~~~(k=1-6),
\label{pJ}
\end{equation}
therefore, equation (\ref{vckm}) can be written as
\begin{equation}
V_{CKM}= U_U^{a \dagger} p_k U_D^{a}.
\label{vckmJ}
\end{equation}
This equation shows that since matrices belonging to each class are related through permutation relations, the CKM matrices obtained from these are also related through permutations. It may be noted that since $p_k$ in equation (\ref{vckmJ}) can only be any of the 6 matrices $p_1-p_6$, one can say that out of all the possible combinations belonging to a particular category, at most 6 will give structurally different CKM matrices. To illustrate this point, we next discuss the combinations belonging to Category 1.

Consider the combinations I-I, i.e., the one  mentioned in Category 1, when both $M_U$ and $M_D$ are matrices from Class I, we first discuss 6 combinations wherein both $M_{U}$ and $M_{D}$ have the same structure, i.e., I$_a$I$_a$, I$_b$I$_b$, etc.. For such combinations, $p_j = p_{j'}$ and therefore, from equation (\ref{pJ}), $p_k = p_j p_{j}^T =I$, hence, all these 6 combinations result into the same CKM matrix, which can be constructed using the expression given in the first row of table \ref{table2}. 
For the remaining 30 combinations wherein $M_U$ and $M_D$ have different structures, it can be checked that 6 combinations I$_a$I$_b$, I$_b$I$_a$, I$_c$I$_f$, I$_d$I$_e$, I$_e$I$_d$ and I$_f$I$_c$ result into same CKM matrix as the product of the corresponding permutations of each of these combinations give $p_k = p_2$, mentioned in equation (\ref{permutations}). Similarly, for the combinations, I$_a$I$_c$, I$_b$I$_e$, I$_c$I$_a$, I$_d$I$_f$, I$_e$I$_b$ and I$_f$I$_d$, again one finds that $p_k = p_3$, while for I$_a$I$_d$, I$_b$I$_f$, I$_c$I$_e$, I$_d$I$_a$, I$_e$I$_c$ and I$_f$I$_b$, we get $p_k = p_4$. Also the combinations I$_a$I$_f$, I$_b$I$_d$, I$_c$I$_b$, I$_d$I$_c$, I$_e$I$_a$ and I$_f$I$_e$, all give $p_k = p_5$ and I$_a$I$_e$, I$_b$I$_c$, I$_c$I$_d$, I$_d$I$_b$, I$_e$I$_f$ and I$_f$I$_a$, all give $p_k = p_6$. Hence, out of the 30 combinations, only 5 of these result in different CKM matrices, which can be expressed in terms of the diagonalizing matrix and permutation matrices as given in table \ref{table2}. Therefore, as mentioned earlier, we find that instead of 36 possible CKM matrices, there exists only 6 structurally different or independent ones for combinations belonging to a particular category.

Similar to the combinations belonging to Category 1, the CKM matrices obtained for other categories are related as well. Out of the 9 combinations for the case II-II, mentioned in Category 2, only 3 are structurally different and lead to 3 independent CKM matrices. Similarly, for each of the combinations I-II and II-I, placed in Category 3 and 4 respectively, only 3 out of 18 combinations result into structurally different CKM matrices as 6 of the combinations result into same permutation $p_k$. Hence, for all the combinations mentioned in the 4 categories, one obtains a total of $6 + 3 + 3 + 3 = 15$ independent CKM matrices.\\
\begin{table}
\centering
\begin{tabular}{|c|c|}

	\hline 
Mass matrices combinations & Resulting CKM matrix  \\  
	\hline
I$_a$I$_a$, I$_b$I$_b$, I$_c$I$_c$, I$_d$I$_d$, I$_e$I$_e$, I$_f$I$_f$	&  $V_{CKM}= U_U^{a \dagger} U_D^{a}$ \\ 
\hline 
I$_a$I$_b$, I$_b$I$_a$, I$_c$I$_f$, I$_d$I$_e$, I$_e$I$_d$, I$_f$I$_c$	&  $V_{CKM}= U_U^{a \dagger} p_2 U_D^{a}$ \\ 
\hline 
I$_a$I$_c$, I$_b$I$_e$, I$_c$I$_a$, I$_d$I$_f$, I$_e$I$_b$, I$_f$I$_d$	&  $V_{CKM}= U_U^{a \dagger} p_3 U_D^{a}$ \\ 
\hline 
I$_a$I$_d$, I$_b$I$_f$, I$_c$I$_e$, I$_d$I$_a$, I$_e$I$_c$, I$_f$I$_b$	&  $V_{CKM}= U_U^{a \dagger} p_4 U_D^{a}$ \\ 
\hline 
I$_a$I$_f$, I$_b$I$_d$, I$_c$I$_b$, I$_d$I$_c$, I$_e$I$_a$, I$_f$I$_e$	&  $V_{CKM}= U_U^{a \dagger} p_5 U_D^{a}$ \\ 
\hline 
I$_a$I$_e$, I$_b$I$_c$, I$_c$I$_d$, I$_d$I$_b$, I$_e$I$_f$, I$_f$I$_a$	&  $V_{CKM}= U_U^{a \dagger} p_6 U_D^{a}$ \\ 
\hline 
\end{tabular}
\vspace{20pt}
\caption{Various possible combinations for I-I and  corresponding expressions to obtain CKM matrices} 
\label{table2} 
\end{table}

To examine the viability of these 15 CKM matrices, we carry out case by case analysis of various combinations. We begin with the possible combinations of I-I given in table \ref{table2}, in order to construct CKM matrices using these expressions, one needs to first obtain the dialganolaizing transformations for I$_a$. The real matrix corresponding to matrix I$_a$ can be expressed as 
\begin{equation}
M_{i}^{r}  =\begin{pmatrix}
  0 & |A_{i}| & 0 \\
  |A_{i}| & D_{i} & |B_{i}| \\
  0 & |B_{i}| & C_{i} \\
\end{pmatrix}
\end{equation}
and $P_{i}$, the phase matrix, is given by
\begin{equation}
    P_{i}=\begin{pmatrix}
      e^{-\text{i$\alpha $}_i} & 0 & 0 \\
      0 & 1 & 0 \\
      0 & 0 & e^{\text{i$\beta $}_i} \\
    \end{pmatrix}
\label{Eq:phase}
\end{equation}
The real matrix $M_{i}^{r}$ can then be diagonalized by the orthogonal transformations $O_{i}$. Using the diagonalizing relations given in equation (\ref{eqIa}), the orthogonal transformation matrix $O_{i}$ turns out to be 
\begin{equation}
O_i\text{  }=\text{     }\left(
\begin{array}{ccc}
	\sqrt{\frac{m_2m_3\left(C_i-m_1\right)}{C_i\left(m_1+m_2\right)\left(m_3-m_1\right)}} & \sqrt{\frac{m_1m_3\left(C_i+m_2\right)}{C_i\left(m_1+m_2\right)\left(m_3+m_2\right)}} & \sqrt{\frac{m_1m_2\left(m_3-C_i\right)}{C_i\left(m_3+m_2\right)\left(m_3-m_1\right)}} \\
	\sqrt{\frac{m_1\left(C_i-m_1\right)}{\left(m_1+m_2\right)\left(m_3-m_1\right)}} & -\sqrt{\frac{m_2\left(C_i+m_2\right)}{\left(m_1+m_2\right)\left(m_3+m_2\right)}} & \sqrt{\frac{m_3\left(m_3-C_i\right)}{\left(m_3-m_1\right)\left(m_2+m_3\right)}} \\
	-\sqrt{\frac{m_1\left(C_i+m_2\right)\left(m_3-C_i\right)}{C_i\left(m_1+m_2\right)\left(m_3-m_1\right)}} & \sqrt{\frac{m_2\left(m_3-C_i\right)\left(C_i-m_1\right)}{C_i\left(m_1+m_2\right)\left(m_3+m_2\right)}} & \sqrt{\frac{m_3\left(C_i-m_1\right)\left(C_i+m_2\right)}{C_i\left(m_3+m_2\right)\left(m_3-m_1\right)}}
\end{array}
\right).
\label{Eq:orthogonal}	
\end{equation}
Using the hierarchy of quark masses, the leading order terms of the diagonalizing matrices are then given by
\begin{equation}
U_U^{a \dagger}= O_U^T P_U \approx \text{     }\left(
\begin{array}{ccc}
 e^{-i \alpha _U} & \sqrt{\frac{m_u}{m_c}} & -e^{i \beta _U} \sqrt{\frac{m_u}{m_t}} \\
 e^{-i \alpha _U} \sqrt{\frac{m_u}{m_c}} & -1 & e^{i \beta _U} \sqrt{\frac{m_c}{m_t}} \\
 e^{-i \alpha _U} \frac{m_c}{m_t}.\sqrt{\frac{m_u}{m_t}} & \sqrt{\frac{m_c}{m_t}} & e^{i \beta _U}
\end{array}
\right),
\label{vuapprox}
\end{equation}
and
\begin{equation}
U_D^{a}= P_D^{\dagger} O_D \approx \text{     }\left(
\begin{array}{ccc}
 e^{i \alpha _D} & e^{i \alpha _D} \sqrt{\frac{m_d}{m_s}} & e^{i \alpha _D} \frac{m_s}{m_b}.\sqrt{\frac{m_d}{m_b}} \\
 \sqrt{\frac{m_d}{m_s}} & -1 & \sqrt{\frac{m_s}{m_b}} \\
 -e^{-i \beta _D} \sqrt{\frac{m_d}{m_b}} & e^{-i \beta _D} \sqrt{\frac{m_s}{m_b}} & e^{-i \beta _D}
\end{array}
\right).
\label{vdapprox}
\end{equation}
It is interesting to note that the diagonal terms of the two matrices in equations (\ref{vuapprox}) and (\ref{vdapprox}) approximate to unity. Also, keeping in mind the hierarchy of the quark masses, it may be noted that the off-diagonal elements are very small in comparison to the diagonal elements. Hence, one can say that these two matrices approximate to the structure of a unit matrix.
Since the product of two unit matrices result into the unit matrix itself, the CKM matrix obtained for the combinations I$_a$I$_a$, I$_b$I$_b$, etc., using the expression given in table (\ref{table2}), has structure that approximates to unit matrix. Such a structure of CKM matrix, where the diagonal elements are nearly unity whereas the off diagonal elements are being much smaller, is referred to as the usual structure as it correseponds to the standard structure of the CKM matrix given by PDG.

For the remaining combinations given in table \ref{table2}, one may note that all these expressions for $V_{CKM}$ contain a permutation matrix. These matrices permute or re-arrange the rows (when pre-multiplying) or columns (when post-multiplying) of the matrix being multiplied to. 
Thus, one of the two matrices, $V_{U}^{a \dag}$ or $V_{D}^a$, gets rearranged such that one of these has the structure of unit matrix corresponding $p_1$, while the other one end up having the structure of one of the remaining 5 permutation matrices. Now, we know that multiplying any matrix with a unit matrix, gives the matrix itself. Hence, the CKM matrices obtained from these two matrices does not have a usual structure of that of a unit matrix, hence, ruling out all of these combinations. Thus, out of 36 possible combinations for I-I, only 6 combinations I$_a$I$_a$, I$_b$I$_b$, etc. lead to the usual structure of CKM matrix. So, it becomes desirable to numerically examine the viability of the CKM matrix obtained for these combinations. Since all these 6 combinations result into exactly same CKM matrix, so, for the purpose of numerical analysis, we consider the combination I$_a$I$_a$, i.e., the Fritzsch like texture 4 zero mass matrices.

Before moving to the numerical work, we first carry analytic analyses for the combinations involving matrices from Class II. As a first step, we find the diagonalizing matrix for the matrices belonging to Class II. The real matrix corresponding to II$_a$ can be expressed as 
\begin{equation}
M_{i}^{r}  =\begin{pmatrix}
  E_{i} & |A_{i}| & 0 \\
  |A_{i}| & 0 & |B_{i}| \\
  0 & |B_{i}| & C_{i} \\
\end{pmatrix}
\label{IIa}
\end{equation}
and $P_{i}$, the phase matrix, is given by
\begin{equation}
    P_{i}=\begin{pmatrix}
      e^{-\text{i$\alpha $}_i} & 0 & 0 \\
      0 & 1 & 0 \\
      0 & 0 & e^{\text{i$\beta $}_i} \\
    \end{pmatrix}
\end{equation}
Using the diagonalizing relations given in equation (\ref{class2}), the orthogonal transformation matrix $O_{i}$ for the real matrix given in equation(\ref{IIa}),turns out to be 
\tiny
\begin{equation}
O_i\text{  }=\text{     }\left(
\begin{array}{ccc}
	\sqrt{\frac{\left(C_i-m_1\right)\left(m_1+m_3-C_i\right))\left(C_i-m_1+m_2\right)}{\left(m_1+m_2\right)\left(m_3-m_1\right)\left(2 C_i-m_1+m_2-m_3\right)}} & \sqrt{\frac{\left(C_i+m_2\right)\left(m_2-m_3+C_i\right))\left(C_i-m_1+m_2\right)}{\left(m_1+m_2\right)\left(m_2+m_3\right)\left(2 C_i-m_1+m_2-m_3\right)}} & \sqrt{\frac{\left(m_3-C_i\right)\left(m_1+m_3-C_i\right))\left(C_i+m_2-m_3\right)}{\left(m_2+m_3\right)\left(m_3-m_1\right)\left(2 C_i-m_1+m_2-m_3\right)}} \\
	\sqrt{\frac{\left(C_i-m_1\right)\left(C_i+m_2-m_3\right)}{\left(m_1+m_2\right)\left(m_3-m_1\right)}} & -\sqrt{\frac{\left(C_i+m_2\right)\left(m_1+m_3-C_i\right)}{\left(m_1+m_2\right)\left(m_2+m_3\right)}} & \sqrt{\frac{\left(C_i-m_1+m_2\right)\left(m_3-C_i\right)}{\left(m_3-m_1\right)\left(m_2+m_3\right)}} \\
	-\sqrt{\frac{\left(m_2+C_i\right)\left(m_3-C_i\right))\left(C_i+m_2-m_3\right)}{\left(m_1+m_2\right)\left(m_3-m_1\right)\left(2 C_i-m_1+m_2-m_3\right)}} & \sqrt{\frac{\left(m_3-C_i\right)\left(C_i-m_1\right)\left(m_1+m_3-C_i\right)}{\left(m_1+m_2\right)\left(m_2+m_3\right))\left(2 C_i-m_1+m_2-m_3\right)}} & \sqrt{\frac{\left(C_i-m_1\right)\left(C_i+m_2\right)\left(C_i-m_1+m_2\right)}{\left(m_2+m_3\right)\left(m_3-m_1\right))\left(2 C_i-m_1+m_2-m_3\right)}}
\end{array}
\right).	
\end{equation}
\normalsize
Using the hierarchy of quark masses, the leading order terms of the diagonalizing matrices are then given by
\begin{equation}
U_U^{a \dagger}= O_U^T P_U \approx \text{     }\left(
\begin{array}{ccc}
 e^{-i \alpha _U} & \sqrt{\frac{m_u}{m_c}} & -e^{i \beta _U} \sqrt{\frac{m_u}{m_t}} \\
 e^{-i \alpha _U} \sqrt{\frac{m_u}{m_c}} & -1 & e^{i \beta _U} \sqrt{\frac{m_c}{m_t}} \\
 e^{-i \alpha _U} \frac{m_c}{m_t}.\sqrt{\frac{m_u}{m_t}} & \sqrt{\frac{m_c}{m_t}} & e^{i \beta _U}
\end{array}
\right),
\label{vuapproxII}
\end{equation}
and
\begin{equation}
U_D^{a}= P_D^{\dagger} O_D \approx \text{     }\left(
\begin{array}{ccc}
 e^{i \alpha _D} & e^{i \alpha _D} \sqrt{\frac{m_d}{m_s}} & e^{i \alpha _D} \frac{m_s}{m_b}.\sqrt{\frac{m_d}{m_b}} \\
 \sqrt{\frac{m_d}{m_s}} & -1 & \sqrt{\frac{m_s}{m_b}} \\
 -e^{-i \beta _D} \sqrt{\frac{m_d}{m_b}} & e^{-i \beta _D} \sqrt{\frac{m_s}{m_b}} & e^{-i \beta _D}
\end{array}
\right).
\label{vdapproxII}
\end{equation}
Keeping in mind the quark mass hierarchy, one can say that these two matrices approximate to the structure of a unit matrix.
 
Hence, for the same reasons mentioned above for the case I-I, only those combinations for II-II, I-II and II-I result into usual CKM matrix for which the set of matrices are obtained from same permutations. In other words, only the combinations II$_a$-II$_a$, II$_b$-II$_b$ etc. belonging to the Category 2, I$_a$-II$_a$, I$_b$-II$_b$ etc. belonging to the Category 3 and II$_a$-I$_a$, II$_b$-I$_b$ etc. for Category 4 give CKM matrix of the usual form and all other combinations are ruled out. Since these 6 combinations for a particular category result into exactly same CKM matrix, hence, for the 4 categories, only 4 combinations, i.e., I$_a$-I$_a$, II$_a$-II$_a$, I$_a$-II$_a$, II$_a$-I$_a$ need to be examined numerically, which involves constructing CKM matrix for these combinations and comparing the matrix so obtained with the recent one given by PDG \cite{pdg20}. 

\section{Inputs}

For the purpose of numerical analysis, we consider the ``current'' quark masses at $M_{Z}$ energy scale \cite{masses}  given by
\begin{eqnarray}
m_{u}=1.23\pm 0.21~{\rm MeV},~~~~m_{d}=2.67\pm 0.19~{\rm MeV}, ~~~~m_{s}=53.16 \pm 4.61~{\rm MeV},~~~~~\nonumber 
\\
m_{c}= 0.620 \pm 0.017 ~{\rm GeV},~~~~ m_{b}=2.839 \pm 0.026~ {\rm GeV},~~~~m_{t}=168.26\pm 0.75~{\rm GeV}. 
\label{masses}
\end{eqnarray} 
The quark mass ratios $\frac{m_u}{m_d}$ and $\frac{m_s}{m_{d}}$, are given by \cite{ratios1}
\begin{equation}
\frac{m_u}{m_d}= 0.56~~~~~{\rm and}~~~~~\frac{m_s}{m_{d}}= 20.2.
\label{massrat}
\end{equation}
In the absence of any information related to the phases associated with the elements of the mass matrices $\phi_{1}=\alpha_{U}-\alpha_{D}$ and $\phi_{2}=\beta_{U}-\beta_{D}$, these have been given full variation from $0^o$ to $360^o$. 
The parameters $D_i$ and $E_i$ are free parameters, however, they have been constrained such that diagonalizing transformations $O_U$ and $O_D$ always remain real. Along with these inputs, we have imposed the recent values of well known CKM matrix elements $V_{ub}$, $V_{cb}$ and CP asymmetry parameter $Sin2\beta$ as per PDG 2020 \cite{pdg20}  as a constraint
\begin{equation}
V_{ub}= (3.82 \pm 0.24) \times 10^{-3},~~~~~V_{cb}= (41.0 \pm 1.4) \times 10^{-3},~~~~~Sin2\beta= 0.699 \pm 0.017.
\label{cons}
\end{equation}
The viability of a set of mass matrices is ensured by examining the compatibility of the CKM matrix constructed from the combination, with the recent one given by Particle Data Group (PDG). CKM matrix obtained by recent global analysis \cite{pdg20} is   
\begin{equation}
    V_{CKM}=\begin{pmatrix}
      0.97390-0.97412 & 0.22602-0.22698 & 0.00352-0.00372 \\
      0.22588-0.22684 & 0.97309-0.97331 & 0.03992-0.04136 \\
      0.00838-0.00877 & 0.03918-0.0406 & 0.999137-0.999196 \\
    \end{pmatrix}.
    \label{pdgmatrix}
\end{equation}
\section{Calculations and results}
We concluded in Section 2 that after ruling out most of the possible texture 4 zero mass matrices, one is left with 4 combinations, namely I$_a$-I$_a$, II$_a$-II$_a$, I$_a$-II$_a$ and II$_a$-I$_a$, which need to be evaluated numerically. In this section, we examine these 4 combinations one by one. Firstly, consider the combination I$_a$-I$_a$, the CKM matrix for this can be obtained from the expression given in first row of table \ref{table2}, where unitary matrices are the product of phase matrix, given in equation (\ref{Eq:phase}) and orthogonal matrix, given in equation (\ref{Eq:orthogonal}). Using the numerical values given in Section 3, the mixing matrix for the combination I$_a$I$_a$ comes out to be,
\begin{equation}
V_{CKM}=\begin{pmatrix}
  0.97389-0.97464 & 0.22371-0.22699 & 0.00358-0.00383 \\
  0.22357-0.22686 & 0.97301-0.97381 & 0.03986-0.04239 \\
  0.00782-0.00875 & 0.03926-0.04168 & 0.999094-0.999198 \\
\end{pmatrix}.
\end{equation}
One can very clearly see the overlap between all the elements of the mixing matrix with the one given by PDG in equation (\ref{pdgmatrix}). This ensures the viability of Fritzsch like texture 4 zero mass matrices  with recent refinements in light quark masses as well as CKM matrix elements. Further, we have also evaluated the Jarlskog's rephasing invariant parameter $J$ and the CP violating phase $\delta$ for this combination, which come out to be 
\begin{equation}
J =(2.72-3.22)\times 10^{-5},~~~\delta=58.15^o-89.85^o.
 \end{equation}
These clearly overlap with their experimentally determined ranges \cite{pdg20} 
\begin{equation}
J =(2.91-3.15)\times 10^{-5},~~~\delta=67.6^o-76.2^o.  
\end{equation}

For the other three combinations, the mixing matrices comes out to be the following 
For II$_a$II$_a$,
\begin{equation}
V_{CKM}=\begin{pmatrix}
  0.9924-0.9990 & 0.0436-0.1229 & 0.0035-0.0041 \\
  0.0431-0.1225 & 0.9845-0.9951 & 0.0801-0.1658 \\
  0.0063-0.0119 & 0.0797-0.1655 & 0.9861-0.9967 \\
\end{pmatrix}.
\end{equation}
For I$_a$II$_a$,
\begin{equation}
V_{CKM}=\begin{pmatrix}
  0.9898-0.9997 & 0.0223-0.1419 & 0.0035-0.0040 \\
  0.0220-0.1418 & 0.9851-0.9964 & 0.0666-0.1063 \\
  0.0053-0.0127 & 0.0662-0.1057 & 0.9943-0.9977 \\
\end{pmatrix}.
\end{equation}
For II$_a$I$_a$,
\begin{equation}
V_{CKM}=\begin{pmatrix}
  0.9742-0.9746 & 0.2237-0.2252 & 0.0035-0.0040 \\
  0.2132-0.2235 & 0.9290-0.9694 & 0.1005-0.3024 \\
  0.0227-0.0679 & 0.0980-0.2947 & 0.9531-0.9949 \\
\end{pmatrix}.
\end{equation}
Clearly, all the 3 CKM matrices, obtained for the combinations involving Class II matrices, show no overlap with the one given by PDG. Hence, all these combinations are ruled out and one can say that only the Fritzsch like texture 4 zero set of mass matrices is a viable texture combination that describes the present day quark mixing data. This is in agreement with our earlier work \cite {unique} where starting with the most general mass matrices, and using the freedom of making WB transformations and imposing the condition of ‘naturalness’ within the texture zero approach, it was found that Fritzsch like texture 4 zero mass matrices can be considered as a unique viable option for the description of quark mixing data.

\begin{figure}
\begin{multicols}{3}
{\hspace*{-45pt}\includegraphics[width=9cm,height=7cm]{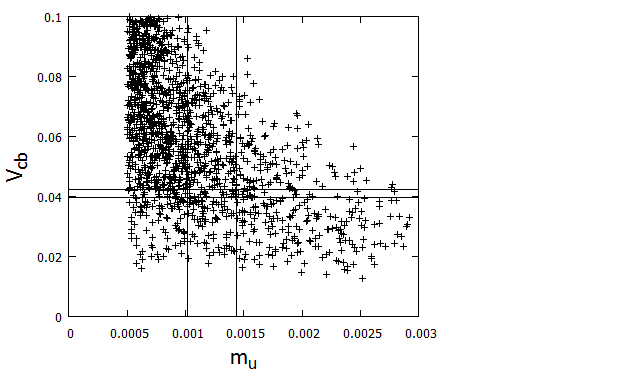}} 
{\hspace*{-20pt}\includegraphics[width=9cm,height=7cm]{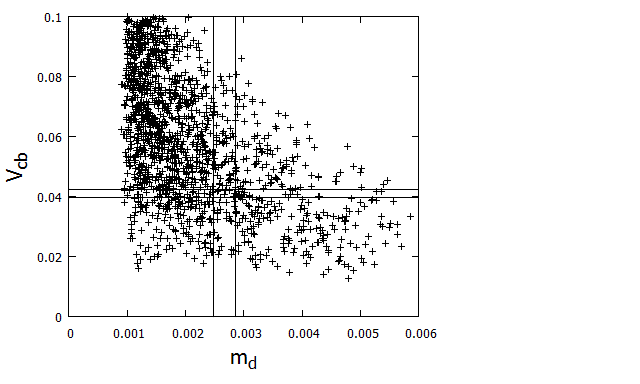}} 
{ \includegraphics[width=9cm,height=7cm]{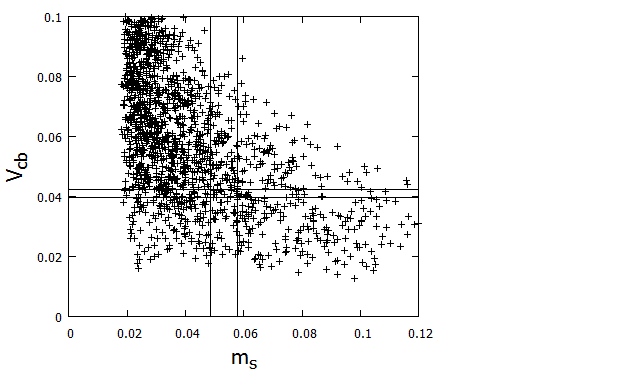}}
\end{multicols} \vspace*{-15pt}
\caption{Allowed range of $V_{cb}$ w.r.t the light quark masses.}
\label{fig:vcb}
\end{figure}

\begin{figure}
\begin{multicols}{3}
{\hspace*{-45pt}\includegraphics[width=9cm,height=7cm]{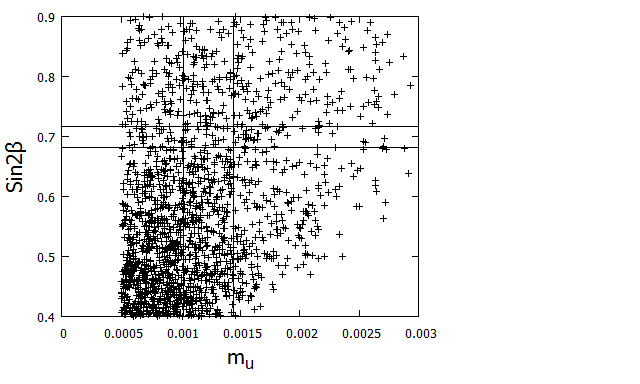}} 
{\hspace*{-20pt}\includegraphics[width=9cm,height=7cm]{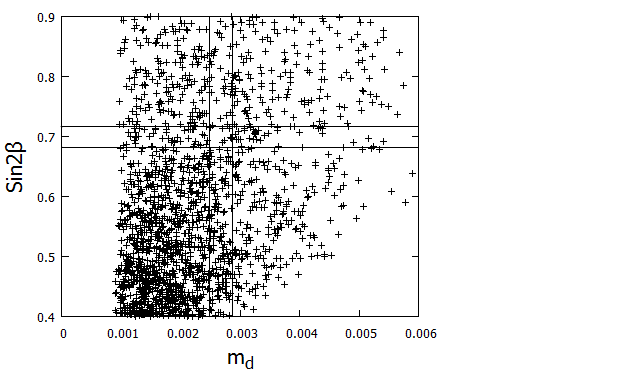}} 
{ \includegraphics[width=9cm,height=7cm]{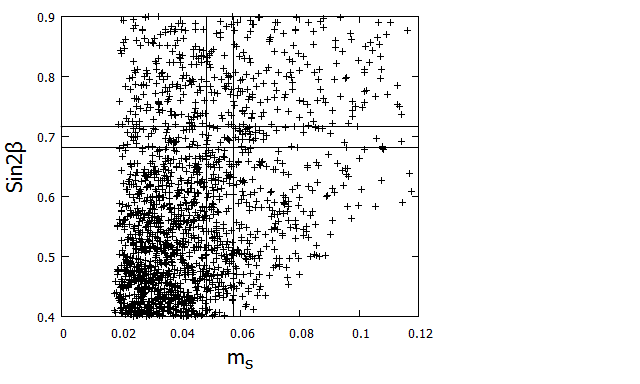}}
\end{multicols} \vspace*{-15pt}
\caption{Allowed range of $Sin2\beta$ w.r.t the light quark masses.}
\label{fig:s2b}
\end{figure}

\begin{figure}
\begin{multicols}{3}
{\hspace*{-45pt}\includegraphics[width=9cm,height=7cm]{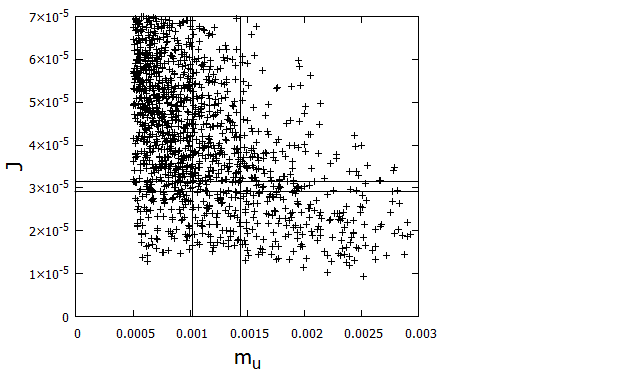}} 
{\hspace*{-20pt}\includegraphics[width=9cm,height=7cm]{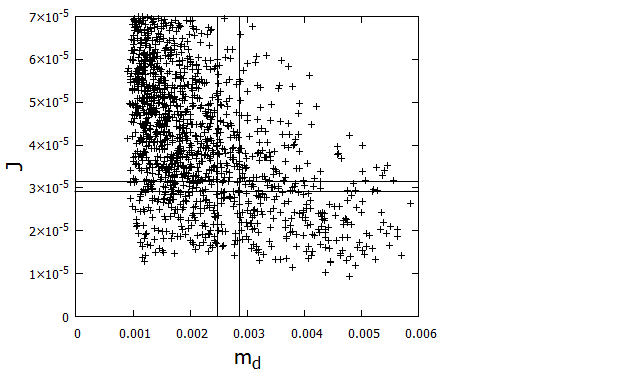}} 
{ \includegraphics[width=9cm,height=7cm]{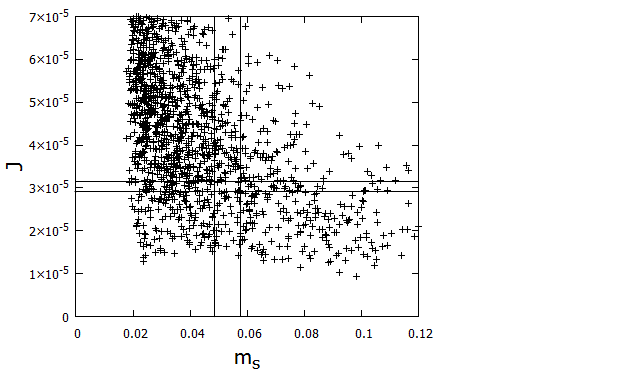}}
\end{multicols} \vspace*{-15pt}
\caption{ Allowed range of $J$ w.r.t the light quark masses.}
\label{fig:J}
\end{figure}

It is to be noted that in the last two decades the quark mass ranges have undergone considerable changes, especially the light quark masses $m_u$, $m_d$, $m_s$. Keeping in mind, attempts to continuous refinements in quark masses, whose mass ranges are found to be fluctuating, it is interesting to examine to what extent Fritzsch like 4 zero mass matrices remain a viable set of mass matrices. To this end, in Figure \ref{fig:vcb}, we have presented the dependence of CKM matrix element $V_{cb}$ w.r.t wide variation of masses $m_u$, $m_d$ and $m_s$. While plotting these graphs, we have considered $m_u$ to have relatively wider range of mass, i.e., from 0.5-3.0 MeV.  We have obtained wider ranges of masses  $m_d$ and $m_s$ using the mass ratios mentioned in equation (\ref{massrat}). The vertical lines in these plots depict the ranges of respective light quark masses given in equation (\ref{masses}), whereas the horizontal lines show the experimental range \cite{pdg20} of the matrix element $V_{cb}$. A look at these plots reveal that even if, in future, the ranges of the light quark masses become considerably wider, the allowed ranges of $V_{cb}$ would clearly overlap with the experimental results, re-emphasizing the earlier conclusion regarding the viability of this combination of mass matrices. We have also shown the variation of CP asymmetry parameter $Sin2\beta$ and Jarlskog's CP invariant parameter $J$ with a wide variation of masses $m_d$ and $m_s$, in Figures \ref{fig:s2b} and \ref{fig:J} respectively. These plots reveal that corresponding to the experimental range of light quark masses, the allowed ranges of $Sin2\beta$ and $J$ obtained here show complete overlap with their experimentally determined ranges. Also, one can clearly see from these figures that even if there are perturbations in the ranges of these parameters or in quark mass ranges, our conclusions remain the same.

\begin{figure}
\subcaptionbox{\label{Fig:s2bdc}}
\centering
{\hspace*{-80pt}\includegraphics[width=7.5cm,height=7cm]{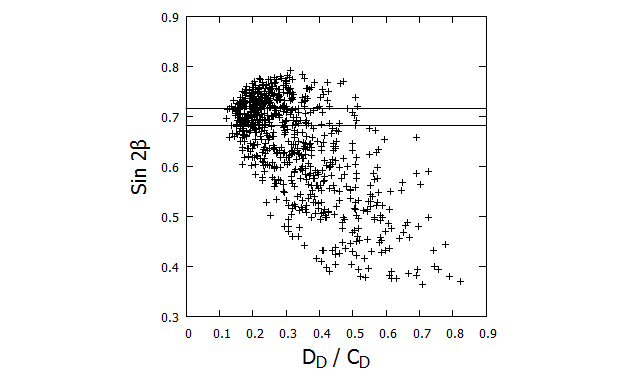}}
\hspace*{-55pt} 
\subcaptionbox{\label{Fig:s2bm}}
{\includegraphics[width=7.5cm,height=7cm]{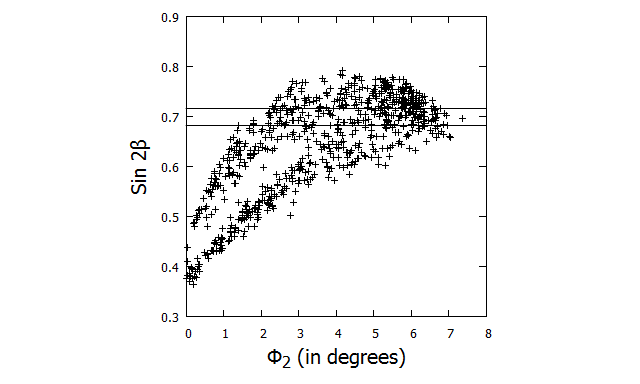}}
\hspace*{-58pt}
\subcaptionbox{\label{Fig:lm}}  
{ \includegraphics[width=7.5cm,height=7cm]{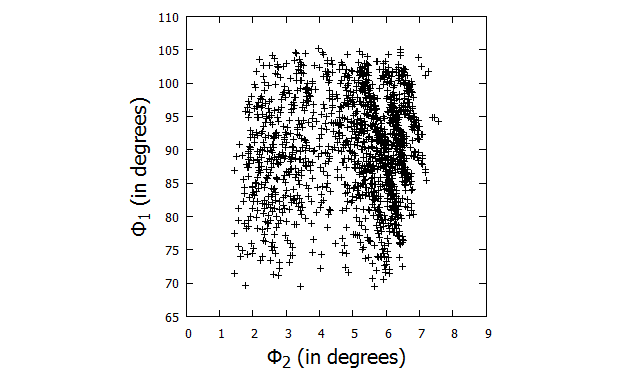}}
\caption{a. Variation of CP violating parameter $Sin2\beta$ versus hierarchy characterizing ratio $D_D/C_D$.\\ b. Variation of CP violating parameter $Sin2\beta$ versus the phase $\phi_2$.\\ c. Allowed ranges of the phases $\phi_1$ and $\phi_2$ of the mass matrix.}
\end{figure}

Keeping in mind that the parameter $Sin2\beta$ provides vital clues to the structural features of texture specific mass matrices, comprising of hierarchy and phases of the elements of the mass matrices \cite{paraspace1, paraspace2, 1so10, sf1, sf2, sf3, sf4, sf5, sf6, sf7} and in view of considerable improvements in the measurement of light quark masses and CKM parameters, including $Sin2\beta$, we have revisited these features for the Fritzsch like texture 4 zero mass matrices. It has been shown in Ref \cite{paraspace1, paraspace2}, that there is limited compatibility of these matrices for strongly hierarchy among the elements of mass matrices, implying $D_i < |B_i| < C_i$,  however, weakly hierarchical matrices, implying, $D_i < |B_i| < C_i$, indicated the compatibility for much broader range of the elements. In the present work, we have investigated the role of hierarchy characterizing ratio $D_D/C_D$ with the present quark mass ranges, given in equation (\ref{masses}). It can be easily noted from graph in Figure \ref{Fig:s2bdc} that only for $D_D/C_D>0.1$, the experimental range of $Sin2\beta$ is reproduced. Hence, it can be clearly concluded that the elements of mass matrices are weakly hierarchical. This is in agreement to the earlier work \cite{sf7}, where it was shown that as one deviates from the strong hierarchy case, only then full range of parameter $\sin2\beta$ is reproduced. 

Regarding the phases having origin in the mass matrices, in ref \cite{paraspace1}, authors have shown that only one of the two phase parameters plays a crucial role in CP violation. However, in another work \cite{paraspace2}, it was shown that both these phases have to be non zero to achieve compatibility of these matrices with quark mixing data. In order to examine whether recent refinements regarding $Sin2\beta$ have implications for these results, in the present work we have plotted the allowed ranges of $Sin2\beta$ with the phase $\phi_2$ in Figure \ref{Fig:s2bm}. It is very clear that present range of $Sin2\beta$ is obtained only when $\phi_2$ is between $1.5^o- 7.0^o$.  In Figure \ref{Fig:lm}, we present the plot  $\phi_1$ versus  $\phi_2$. Interestingly, the present refined inputs limit the allowed ranges of the two phases to $\phi_1 \sim 70^o-105^o$  and $\phi_2 \sim 1.5^o-7.0^o$. 

\section{Summary and conclusions}
In view of good deal of refinements in the measurements of small quark masses $m_u$, $m_d$ and $m_s$ as well as in the CKM matrix elements, we have carried out an extensive analysis of all possible texture 4 zero quark mass matrices. Interestingly, many of these combinations can be ruled out analytically. For the remaining, corresponding CKM matrices have been constructed and compared with the latest mixing data. One finds that out of all these possibilities, only Fritzsch like texture 4 zero mass matrices are compatible with recent results emerging from global fits. Further, we have found that these conclusions would remain largely valid even if, in future, there are changes in the ranges of the light quark masses. We have also examined the implications of recent precision measurements on the structural features of Fritzsch-like texture 4 zero quark mass matrices. We find, the (2,2), (2,3) and (3,3) elements of the up or down type quark mass matrix have a relatively weak hierarchy, although their magnitudes are considerably larger than the magnitude of the (1,2) element. Further, in view of the more precise information regarding CP violating parameters, the ranges of both the phases $\phi_1$ and $\phi_2$, having their origin in the mass matrices, have been found. We find that both the phase parameters plays a crucial role in CP violation. Such structural features might have important implications for model building of fermion mass matrices.
\\
\\
{\bf Acknowledgements} \\ The authors would like to thank Chairman, Department of Physics, P.U., for providing facilities to work.

\end{document}